\documentclass[journal]{IEEEtran}

\usepackage{cite}
\usepackage{amsmath,amssymb,amsfonts}
\usepackage{graphicx}
\usepackage[caption=false,font=footnotesize]{subfig}
\usepackage{booktabs}
\usepackage{multirow}
\usepackage{array}
\usepackage{tikz}
\usetikzlibrary{positioning,arrows.meta,fit}
\usepackage{url}
\usepackage[hidelinks]{hyperref}

\begin{document}
\bstctlcite{IEEEexample:BSTcontrol}

\title{Window Size Matters: Benchmarking Voice Activity Detectors in Noise}

\author{Alok~Kumar, Mohit~Srivastava, Anupam~Purwar$^{*}$, Kuldeep~Kumar~Nigam,\\ Yash~Anchaliya, Pranav~Wadhwa, Debrup~Paul\thanks{$^{*}$Corresponding author: Anupam Purwar (e-mail: \href{mailto:anupam@onebyzero.ai}{anupam@onebyzero.ai}).}\\
{\normalsize OneByZero AI}}

\markboth{}%
{Kumar \MakeLowercase{\textit{et al.}}: Window Size Matters}

\maketitle

\begin{abstract}
Voice activity detection (VAD) sits in front of almost every streaming speech recognizer, yet the choice of detector is often made by feel rather than by measurement. We report a frame-level offline benchmark of three detectors on the \texttt{test-clean} split of LibriVAD-small: the neural TEN VAD, the neural Silero VAD and a simple RMS energy baseline. The evaluation covers 2,214 noisy utterances (about 7.1 hours) built from LibriSpeech, nine noise families and six SNR levels from $-5$ to 20~dB. At a 16~ms hop, TEN VAD reaches a ROC AUC of 0.917, an average precision of 0.968 and an F1 of 0.913. A sweep over 20, 30, 40 and 50~ms windows shows TEN to be almost insensitive to frame length (AUC 0.916 to 0.919), while RMS stays near 0.75 to 0.77. Silero matches TEN only near its native 32~ms input size and drops to 0.63 at 20~ms. We hypothesize that this is likely caused by our fixed-frame protocol rather than by the model itself. Zero padding, sub-chunk averaging, unreset model state and skipped post-processing are the likely factors, but we have not ablated them. In a per-condition breakdown of TEN, babble noise at low SNR stays close to chance. We also show that F1 at a fixed 0.5 threshold is a weak metric on this data, because a trivial all-speech detector already scores 0.854. All settings needed to reproduce the results are given in the paper.
\end{abstract}


\section{Introduction}
\IEEEPARstart{V}{oice} activity detection decides, for each short frame of audio, whether someone is speaking. In a real-time speech recognition stack it controls when audio is sent to the recognizer, when an utterance is finalized and how much compute is spent on silence. A detector that misses speech cuts words. A detector that fires on noise wastes GPU time and can trigger early or late endpoints.

VAD matters more now that cascaded voice agents are moving into production. In i-LAVA, Purwar and Choudhary \cite{purwar2025ilava} profiled a low-latency voice-to-voice pipeline built from ASR, dialog management and TTS. They found that TTS dominated the real-time factor. That study treated the speech front end as given. FOCAL \cite{purwar2026focal} proposed a benchmark for multi-modal agents that tracks how errors move from one stage of a cascade to the next. MM-tau-p$^2$ \cite{purwar2026mmtau} extended agent evaluation to dual-control settings, with persona adaptation and twelve new metrics. A study of LLM judges for voice agents \cite{purwar2026judges} showed that end-to-end scores also depend on pipeline factors such as streaming and ASR error propagation. On the generation side, Purwar and Choudhary \cite{purwar2026tts} found that LoRA fine-tuning of an LLM-based TTS model helps only when the training data are acoustically diverse. A common thread runs through these papers. System-level metrics can show that an agent failed. They cannot easily show whether the failure started at the very first stage, where the VAD decides what audio reaches the recognizer. A clipped word or a missed onset there becomes an ASR error, then a reasoning error, and finally a poor judge score. This paper isolates that first stage and measures it directly.

Our team runs a real-time ASR service built on NVIDIA Parakeet models \cite{rekesh2023fastconformer} served through NeMo \cite{kuchaiev2019nemo}. Two front ends exist in that service. One uses a plain RMS energy gate. The other uses TEN VAD \cite{tenvad2025}, a small neural detector. Silero VAD \cite{silero2021} is a popular alternative that we wanted to compare against before deciding on a default. We could not find a public, reproducible comparison of these three detectors on a controlled noisy English benchmark, so we built one.

We use LibriVAD \cite{stylianou2025librivad}, a recent open dataset that mixes LibriSpeech \cite{panayotov2015librispeech} with real and synthetic noise and provides sample-level speech labels from forced alignment. LibriVAD gives exact control over noise type and SNR, which makes it a good fit for this question.

The contributions of this paper are practical:
\begin{enumerate}
\item A frame-level offline benchmark of TEN, Silero and RMS on the full LibriVAD-small \texttt{test-clean} split (2,214 files, 1.6~M frames at 16~ms).
\item A window-size sweep from 20 to 50~ms that exposes very different sensitivity across detectors.
\item A discussion of likely reasons why Silero looks fragile under fixed framing, and of why F1 at a fixed threshold can mislead on speech-heavy data.
\item A stratified view of TEN by noise family and SNR, showing that babble is its main weakness.
\end{enumerate}

We stress that this is an offline study. It measures how well each detector separates speech from non-speech frames. It does not measure streaming latency or endpointing behavior, which we leave to future work.

\section{Related Work}

\subsection{Classical detectors}
Early VADs in telephony used hand-tuned features such as energy, zero crossing rate and spectral shape, as in ITU-T G.729 Annex~B \cite{benyassine1997g729b}. Sohn \textit{et al.} \cite{sohn1999statistical} framed VAD as a likelihood ratio test under Gaussian models of speech and noise spectra. Ram\'irez \textit{et al.} \cite{ramirez2004ltsd} added long-term spectral divergence to improve robustness at low SNR. The WebRTC detector \cite{webrtcvad}, built on Gaussian mixture models, is still widely used because it is tiny and fast. rVAD \cite{tan2020rvad} is an unsupervised, segment-based method that combines a priori SNR weighting with pitch information and holds up well in noise. Energy gating, which we use as a baseline, is the simplest member of this family and is still common in production because it costs almost nothing.

\subsection{Neural detectors}
Hughes and Mierle \cite{hughes2013recurrent} showed that recurrent networks outperform GMM-based VAD on noisy speech. Zhang and Wang \cite{zhang2016boosting} proposed boosted DNNs that use multi-resolution context. Zazo \textit{et al.} \cite{zazo2016feature} trained CLDNNs \cite{sainath2015cldnn} directly on raw waveforms. Thomas \textit{et al.} \cite{thomas2014analyzing} studied CNNs under mismatched acoustic conditions. MarbleNet \cite{jia2021marblenet} is a compact 1D separable convolution model shipped with NeMo. Silero VAD \cite{silero2021} and TEN VAD \cite{tenvad2025} are two recent pretrained detectors aimed at on-device and streaming use. Both expose a per-frame speech probability, which makes them easy to score with threshold-free metrics.

\subsection{Datasets}
Many VAD papers report results on private or task-specific data, which makes comparison hard. AVA-Speech \cite{chaudhuri2018ava} offered dense labels on movie audio. VOiCES \cite{richey2018voices} provides far-field recordings in real rooms. LibriVAD \cite{stylianou2025librivad} takes a different route. It synthesizes noisy mixtures from LibriSpeech at controlled SNR and silence-to-speech ratio, uses forced alignments for labels \cite{mcauliffe2017mfa}, and ships three sizes (about 15~GB, 150~GB and 1.5~TB). The LibriVAD authors benchmark several trained models, including a Vision Transformer. We complement that work by evaluating off-the-shelf pretrained detectors that practitioners actually deploy, with no training on LibriVAD at all.

\section{Dataset}

\subsection{Generation}
\begin{figure}[!t]
\centering
\begin{tikzpicture}[
  node distance=3mm and 4mm,
  box/.style={draw, rounded corners=2pt, align=center, font=\scriptsize, minimum height=6mm, inner sep=2pt, text width=17mm},
  arr/.style={-{Latex[length=1.6mm]}, thick}]
\node[box] (ls) {LibriSpeech\\test-clean};
\node[box, below=of ls] (nz) {9 noise\\families};
\node[box, below=of nz] (al) {Forced\\alignments};
\node[box, right=of nz, text width=15mm] (mix) {Mix at SNR\\$-5$ to 20 dB};
\node[box, right=of mix, text width=15mm] (wav) {Noisy WAV\\16 kHz};
\node[box, text width=15mm] (lab) at (al -| wav) {Sample labels\\(.npy)};
\node[box, right=of wav, text width=14mm] (frm) {Framing\\hop $=H$};
\node[box, text width=14mm] (eng) at (al -| frm) {Detector\\score $s_k$};
\node[box, below=of eng, text width=14mm] (met) {AUC, AP,\\F1@0.5};
\draw[arr] (ls) -| (mix);
\draw[arr] (nz) -- (mix);
\draw[arr] (mix) -- (wav);
\draw[arr] (al) -- (lab);
\draw[arr] (wav) -- (frm);
\draw[arr] (frm) -- (eng);
\draw[arr] (eng) -- (met);
\draw[arr] (lab) |- (met);
\end{tikzpicture}
\caption{Data generation and evaluation pipeline. Clean speech is mixed with noise at six SNR levels. Labels come from forced alignment of the clean speech. Each detector scores non-overlapping frames of size $H$.}
\label{fig:pipeline}
\end{figure}
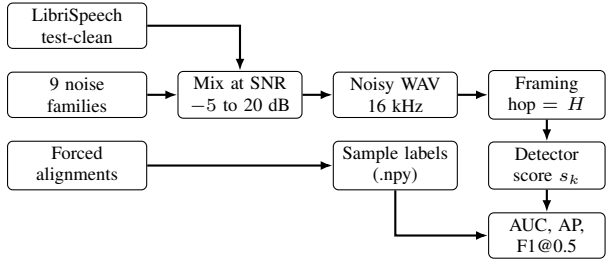

LibriVAD is not hand collected. The open generator \cite{stylianou2025librivad,librivadrepo} takes clean LibriSpeech utterances, mixes each one with a noise recording at a fixed SNR, and writes a per-sample label mask (0 for silence, 1 for speech) derived from forced alignment of the clean signal. Fig.~\ref{fig:pipeline} summarizes the pipeline we used.

Two utterance styles are produced. The \textit{LibriSpeech} style keeps single utterances. The \textit{LibriSpeechConcat} style concatenates utterances, which changes the amount and placement of silence. Noise comes from nine families: Babble, City, Domestic, Nature, Office, Public, speech-shaped noise (SSN), Street and Transport. Each family is mixed at SNRs of $-5$, 0, 5, 10, 15 and 20~dB.

\subsection{Subset used in this study}
The data are also distributed through Hugging Face \cite{librivadhf}. We generated the \textit{small} variant with the upstream script and evaluated only the \texttt{test-clean} split. Table~\ref{tab:data} lists the details. The \texttt{train-clean-100} and \texttt{dev-clean} mixes were generated as a side effect but not used, since no detector was trained.

\begin{table}[!t]
\caption{Evaluation Data}
\label{tab:data}
\centering
\footnotesize
\begin{tabular}{@{}ll@{}}
\toprule
Item & Value \\
\midrule
Benchmark & LibriVAD \cite{stylianou2025librivad}, small variant \\
Source speech & LibriSpeech \cite{panayotov2015librispeech}, English \\
Split & \texttt{test-clean} \\
Files & 2,214 noisy WAVs (0 skipped) \\
Duration & about 7.1 h \\
Sample rate & 16 kHz, 16-bit PCM \\
Labels & per sample, 0 = silence, 1 = speech \\
Noise families & 9 \\
SNR levels & $-5$, 0, 5, 10, 15, 20 dB \\
Speech prior (16 ms frames) & 0.745 \\
Full generated set & about 27,648 WAVs, about 15 GB \\
\bottomrule
\end{tabular}
\end{table}

Note that the speech prior is high. About three quarters of all frames are labeled speech. This matters for how metrics should be read, as we discuss in Section~\ref{sec:metrics_disc}.

\section{Method}

\subsection{Framing and labels}
Let $x[n]$ be a 16~kHz signal and $y[n]\in\{0,1\}$ its sample-level label. For a window of $W$ ms the hop is
\begin{equation}
H = \operatorname{round}\!\left(\frac{16000\cdot W}{1000}\right),
\end{equation}
so $W = 20, 30, 40, 50$~ms gives $H = 320, 480, 640, 800$ samples. Frames do not overlap. Frame $k$ covers samples $kH$ to $(k+1)H-1$, and its reference label $\ell_k$ is obtained by aggregating $y[n]$ over that range. Each detector returns a continuous score $s_k\in[0,1]$. We pool all frames from all files and compute metrics on the pooled set. This is batch scoring, not a streaming state machine. No hangover, smoothing or minimum duration rules are applied.

\subsection{Detectors}

\textit{TEN VAD.} We use the \texttt{ten\_vad} Python package \cite{tenvad2025}, installed from PyPI, with \texttt{TenVad(hop\_size=H, threshold=0.5)} and take the raw speech probability returned by \texttt{process()} for each frame. This is the same library used by our TEN-based Parakeet server. The fixed-hop experiment uses $H=256$ (16~ms), which matches our production session settings. Note that the TEN documentation lists only 160 and 256 samples (10 and 16~ms) as optimized hop sizes. The window sweep runs it at 320 to 800 samples, which the library accepts but which lies outside the documented settings.

\textit{RMS energy.} For each frame we compute
\begin{equation}
r_k = \sqrt{\frac{1}{H}\sum_{n=kH}^{(k+1)H-1} \left(\frac{x[n]}{32768}\right)^2},\qquad
d_k = 20\log_{10}(r_k+\epsilon),
\end{equation}
and map the dBFS value to a score with
\begin{equation}
s_k = \operatorname{clip}\!\left(\frac{d_k + 100}{100},\,0,\,1\right).
\end{equation}
This gives a monotone score suitable for ROC analysis. Our live RMS server uses the same energy idea with fixed endpointing thresholds.

\textit{Silero VAD.} We load the pretrained model with \texttt{load\_silero\_vad()} from the \texttt{silero-vad} package \cite{silero2021}, also installed from PyPI. The installed version rejects inputs shorter than 512 samples (32~ms). For $H\le512$ we zero-pad the frame to 512 samples before scoring. For $H>512$ we split the frame into 512-sample sub-chunks and average their probabilities. One model instance is created per run and its internal state is never reset, so the recurrent state carries over from frame to frame and from one file to the next. Silero is not deployed in our servers. It is included as a reference point.

No detector was trained or fine-tuned on LibriVAD.

Table~\ref{tab:detectors} summarizes how each detector is driven in our harness. The most important row is the input constraint, since it decides whether a given window can be scored without any adaptation.

\begin{table}[!t]
\caption{Detector Configuration in the Evaluation Harness}
\label{tab:detectors}
\centering
\footnotesize
\setlength{\tabcolsep}{4pt}
\begin{tabular}{@{}>{\raggedright\arraybackslash}p{17mm}>{\raggedright\arraybackslash}p{18mm}>{\raggedright\arraybackslash}p{19mm}>{\raggedright\arraybackslash}p{17mm}@{}}
\toprule
 & TEN VAD & Silero VAD & RMS \\
\midrule
Type & Neural, pretrained & Neural, pretrained & Signal energy \\
Score & Speech prob. & Speech prob. & Mapped dBFS \\
Input size & Any hop $H$ & $\ge$512 samples & Any hop $H$ \\
Documented hop & 160 or 256 & 512 & n/a \\
Hops run here & 256 to 800\textsuperscript{a} & 320 to 800 & 320 to 800 \\
State reset & n/a & Never & n/a \\
Short frames & Native & Zero-padded & Native \\
Long frames & Native & Mean of 512-sample chunks & Native \\
Trained here & No & No & n/a \\
In our servers & Yes & No & Yes \\
\bottomrule
\multicolumn{4}{@{}p{75mm}@{}}{\scriptsize \textsuperscript{a}Hops of 320 to 800 samples are outside the sizes TEN documents as optimized.}
\end{tabular}
\end{table}

\subsection{Metrics}
We report three frame-level metrics on the pooled frames: the area under the ROC curve (AUC) \cite{fawcett2006roc}, average precision (AP), which summarizes the precision-recall curve \cite{davis2006relationship}, and F1 at a fixed threshold of 0.5. For a threshold $\tau$ the true and false positive rates are
\begin{equation}
\begin{aligned}
\mathrm{TPR}(\tau)&=\frac{\sum_k \mathbf{1}[s_k\ge\tau]\,\ell_k}{\sum_k \ell_k},\\
\mathrm{FPR}(\tau)&=\frac{\sum_k \mathbf{1}[s_k\ge\tau]\,(1-\ell_k)}{\sum_k (1-\ell_k)}.
\end{aligned}
\end{equation}
AUC is the area under the curve traced by $(\mathrm{FPR},\mathrm{TPR})$ as $\tau$ varies. It equals the probability that a random speech frame receives a higher score than a random non-speech frame, so 0.5 is chance regardless of class balance. AP is computed as
\begin{equation}
\mathrm{AP}=\sum_{i}(R_i-R_{i-1})\,P_i,
\end{equation}
where $P_i$ and $R_i$ are precision and recall at the $i$-th distinct threshold. Unlike AUC, the chance level of AP equals the positive class prior. AUC and AP are threshold free and measure ranking quality. F1@0.5 reflects a deployable operating point but depends on calibration. For the window sweep we also compute AUC per file and show its distribution. For TEN at 16~ms we further break results down by noise family and SNR.

\subsection{Experiments}
\textit{Experiment A} runs TEN alone at $H=256$ on all 2,214 files and stores stratified metrics. \textit{Experiment B} runs TEN, RMS and Silero at $W\in\{20,30,40,50\}$~ms on the same files. Both experiments use the same data loader, framing code and metric implementation, so their numbers are directly comparable.

\section{Results}

\subsection{Experiment A: TEN VAD at 16 ms}
Table~\ref{tab:tenfixed} gives the headline numbers. Across 1,608,606 frames TEN reaches an AUC of 0.917 and an AP of 0.968. As a sanity check, $1{,}608{,}606\times 256 / 16000 \approx 25{,}738$~s, or 7.15~h, which matches the dataset duration.

\begin{table}[!t]
\caption{TEN VAD at Hop 256 (16 ms), LibriVAD-small \texttt{test-clean}}
\label{tab:tenfixed}
\centering
\footnotesize
\begin{tabular}{@{}lr@{}}
\toprule
Metric & Value \\
\midrule
Files scored & 2,214 \\
Frames & 1,608,606 \\
Speech prior & 0.745 \\
ROC AUC & 0.917 \\
Average precision & 0.968 \\
F1 @ 0.5 & 0.913 \\
\bottomrule
\end{tabular}
\end{table}

The stratified results show a clear dependence on SNR. Table~\ref{tab:babble} lists the babble strata for the LibriSpeech style. At $-5$~dB, TEN is essentially at chance (AUC 0.509). It climbs steadily and reaches 0.916 at 20~dB. Most non-babble strata at high SNR exceed 0.95 AUC. For TEN, babble is the hardest family by a wide margin. This is plausible, since the interfering signal is itself speech and offers no simple spectral cue to reject it. We did not stratify RMS or Silero, so we cannot say how they behave in this condition.

\begin{table}[!t]
\caption{TEN VAD on Babble Noise by SNR (LibriSpeech Style, 16 ms)}
\label{tab:babble}
\centering
\footnotesize
\begin{tabular}{@{}lcc@{}}
\toprule
SNR (dB) & ROC AUC & AP \\
\midrule
$-5$ & 0.509 & 0.834 \\
0 & 0.555 & 0.859 \\
5 & 0.657 & 0.907 \\
10 & 0.766 & 0.945 \\
15 & 0.854 & 0.969 \\
20 & 0.916 & 0.983 \\
\bottomrule
\end{tabular}
\end{table}

Note how AP stays above 0.83 even when AUC is near 0.5. This is the speech prior at work. With about 75\% speech frames, a random ranking already yields AP close to the prior.

\subsection{Experiment B: window sweep}
\begin{figure*}[!t]
\centering
\includegraphics[width=\textwidth]{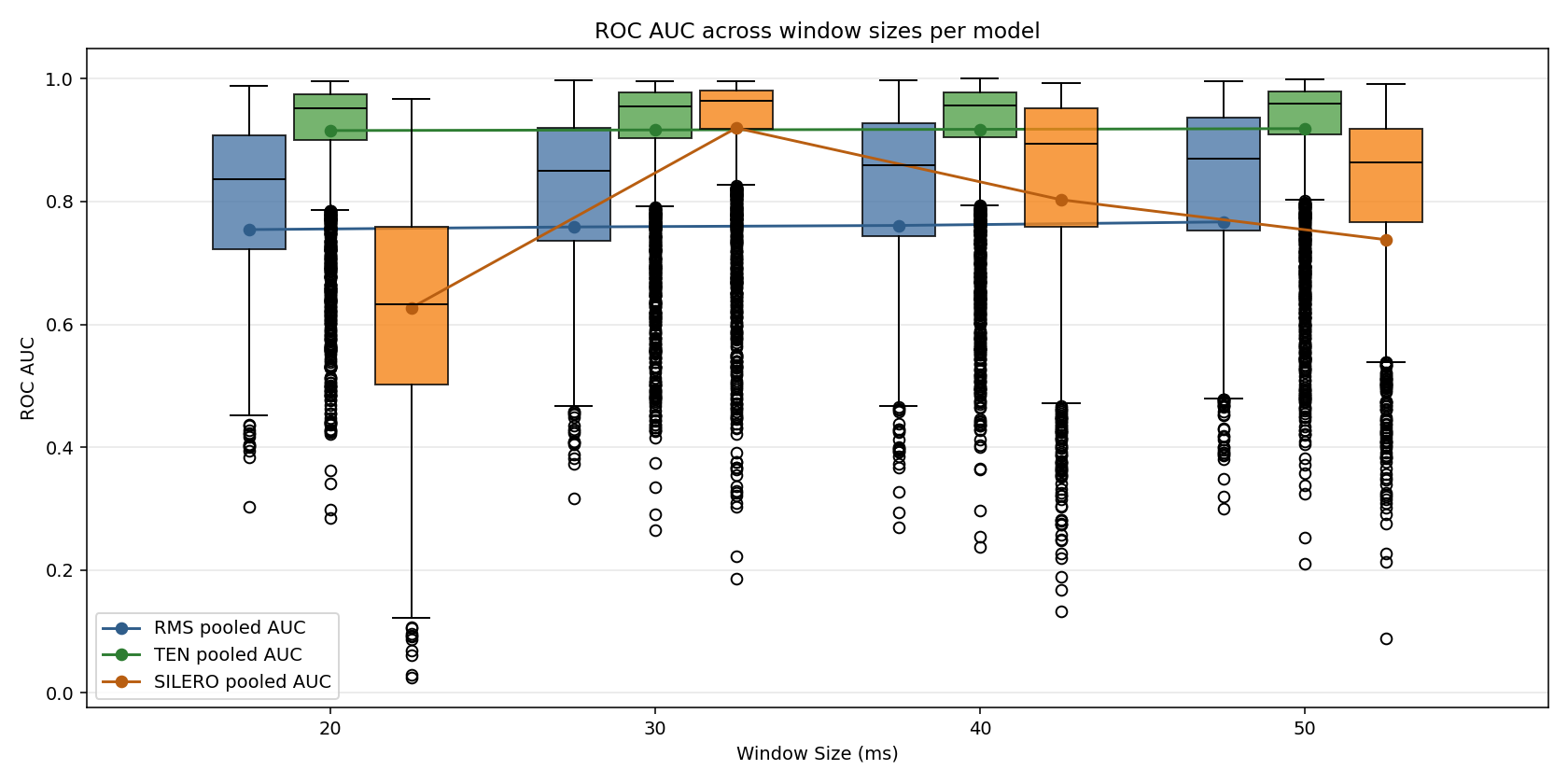}
\caption{Per-file ROC AUC by detector and window size (boxes) with pooled AUC overlaid (lines). TEN is in green, RMS in blue and Silero in orange. TEN stays near 0.92 pooled AUC at every window. Silero peaks at 30 ms.}
\label{fig:box}
\end{figure*}

\begin{figure*}[!t]
\centering
\subfloat[TEN]{\includegraphics[width=0.42\textwidth]{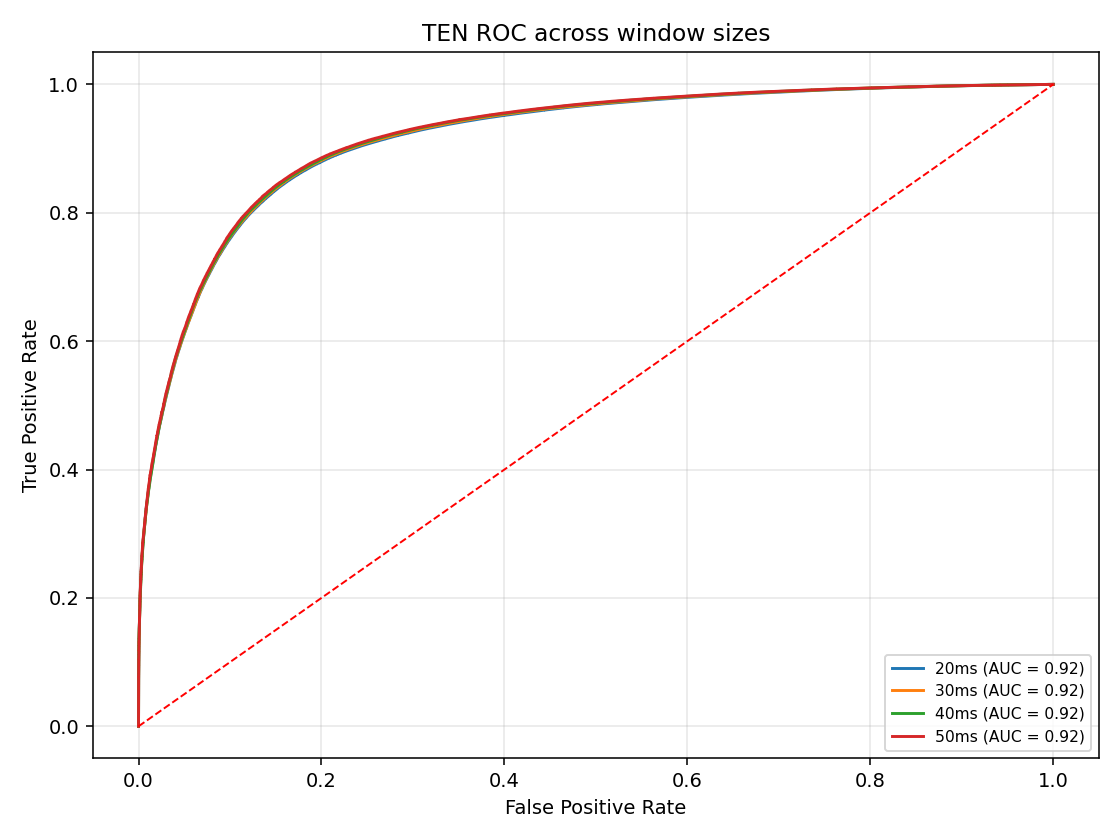}\label{fig:roc_ten}}
\hfil
\subfloat[RMS]{\includegraphics[width=0.42\textwidth]{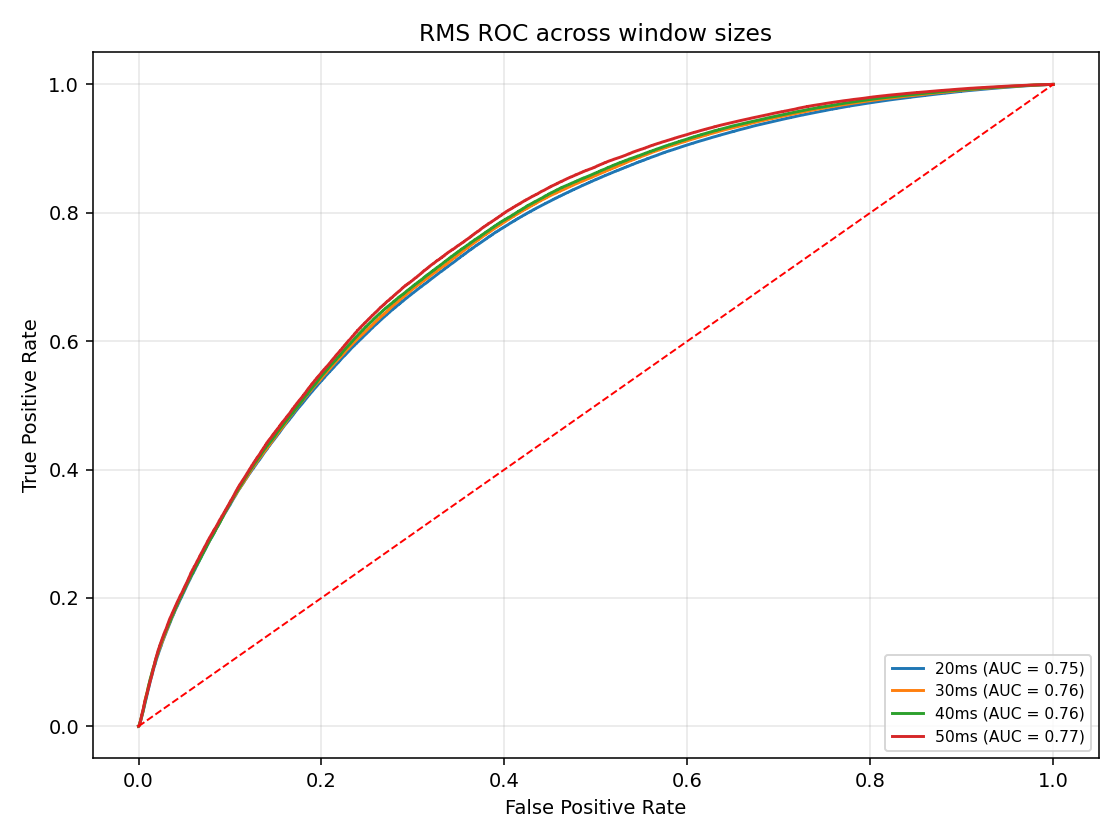}\label{fig:roc_rms}}
\\[2pt]
\subfloat[Silero]{\includegraphics[width=0.42\textwidth]{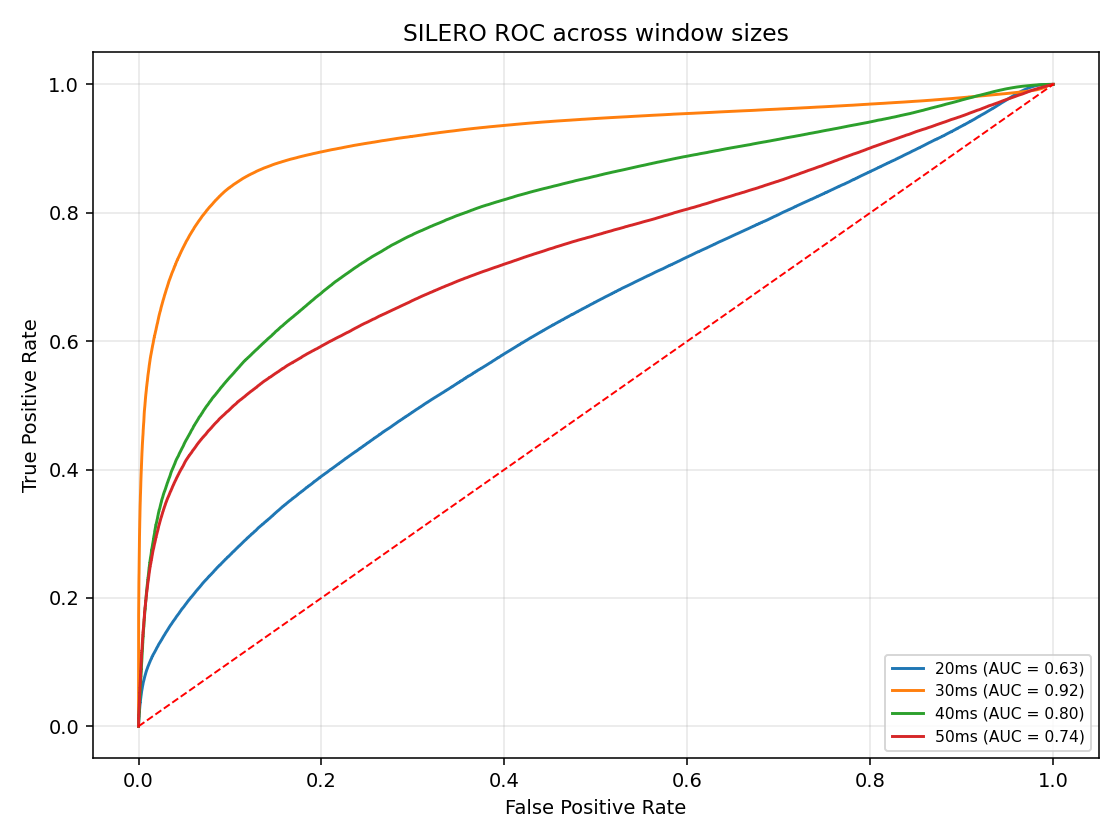}\label{fig:roc_silero}}
\caption{Pooled frame-level ROC curves for window sizes of 20, 30, 40 and 50 ms. The dashed diagonal is chance. TEN curves overlap almost exactly. Silero curves separate widely.}
\label{fig:roc}
\end{figure*}

Table~\ref{tab:sweep} reports pooled metrics for all detectors and window sizes. Fig.~\ref{fig:box} shows per-file AUC distributions together with the pooled AUC, and Fig.~\ref{fig:roc} shows the pooled ROC curves.

\begin{table}[!t]
\caption{Window Sweep on 2,214 Files (Pooled Frame-Level Metrics)}
\label{tab:sweep}
\centering
\footnotesize
\setlength{\tabcolsep}{5pt}
\begin{tabular}{@{}llrccc@{}}
\toprule
Detector & $W$ (ms) & Frames & AUC & AP & F1@0.5 \\
\midrule
\multirow{4}{*}{TEN}
 & 20 & 1,286,712 & 0.916 & 0.968 & 0.913 \\
 & 30 & 857,520 & 0.917 & 0.968 & 0.913 \\
 & 40 & 642,870 & 0.918 & 0.969 & 0.914 \\
 & 50 & 513,918 & \textbf{0.919} & 0.969 & \textbf{0.915} \\
\midrule
\multirow{4}{*}{RMS}
 & 20 & 1,286,712 & 0.754 & 0.884 & 0.856 \\
 & 30 & 857,520 & 0.759 & 0.884 & 0.854 \\
 & 40 & 642,870 & 0.761 & 0.887 & 0.856 \\
 & 50 & 513,918 & 0.767 & 0.888 & 0.855 \\
\midrule
\multirow{4}{*}{Silero}
 & 20 & 1,286,712 & 0.627 & 0.842 & 0.353 \\
 & 30 & 857,520 & \textbf{0.920} & \textbf{0.974} & 0.910 \\
 & 40 & 642,870 & 0.803 & 0.926 & 0.283 \\
 & 50 & 513,918 & 0.738 & 0.903 & 0.533 \\
\bottomrule
\end{tabular}
\end{table}

\textit{TEN} is remarkably stable. AUC moves from 0.916 to 0.919 and AP from 0.968 to 0.969 across the whole sweep. The four ROC curves in Fig.~\ref{fig:roc_ten} are indistinguishable. Longer windows help a little, which fits the intuition that more context per decision reduces noise in the score.

\textit{RMS} is also stable but much weaker, with pooled AUC between 0.754 and 0.767. Its per-file median AUC in Fig.~\ref{fig:box} is noticeably higher, at roughly 0.84 to 0.87. We think the gap comes from pooling. The dBFS mapping is absolute, so a frame of noise in a loud file can score higher than a frame of speech in a quiet file. Within one file, energy ranks frames reasonably well. Across files with different SNRs and noise levels it does not.

A useful way to compare TEN and RMS is the ranking error $1-\mathrm{AUC}$. At 20~ms it is 0.084 for TEN and 0.246 for RMS, so TEN removes about two thirds of the ranking errors made by the energy gate. The ratio is almost the same at 50~ms (0.081 versus 0.233).

\textit{Silero} behaves very differently. At 30~ms it is the best detector in the table by a small margin (AUC 0.920, AP 0.974). At 20~ms it falls to 0.627, and at 40 and 50~ms it reaches only 0.803 and 0.738. Its F1@0.5 swings from 0.910 down to 0.283.

\section{Discussion}

\subsection{Why TEN does better}
The two neural detectors are built for different goals. Table~\ref{tab:tenvssilero} summarizes their public documentation. Silero is a general-purpose model. Its authors report training on very large corpora covering more than 6,000 languages, with varied noise and recording quality \cite{silero2021}. It processes fixed chunks of 30~ms or more, and its main entry point, \texttt{get\_speech\_timestamps()}, wraps the raw probabilities in its own thresholding and duration rules. TEN is narrower. It targets conversational agents and stresses fast transitions between speech and non-speech, including short pauses \cite{tenvad2025}. It exposes a per-frame probability at a hop chosen by the user.

\begin{table}[!t]
\caption{TEN VAD and Silero VAD as Documented by Their Authors}
\label{tab:tenvssilero}
\centering
\footnotesize
\setlength{\tabcolsep}{4pt}
\begin{tabular}{@{}>{\raggedright\arraybackslash}p{17mm}>{\raggedright\arraybackslash}p{29mm}>{\raggedright\arraybackslash}p{29mm}@{}}
\toprule
 & TEN VAD \cite{tenvad2025} & Silero VAD \cite{silero2021} \\
\midrule
Design goal & Conversational agents, fast speech and pause transitions & General purpose, many domains \\
Training data & Not disclosed & Large corpora, 6,000+ languages, varied noise \\
Input unit & Hop of 160 or 256 samples & Fixed chunk of 30~ms or more \\
Sample rate & 16 kHz & 8 and 16 kHz \\
Explicit voicing cue & Pitch estimator in source & None documented \\
Intended output & Per-frame probability & Speech timestamps after post-processing \\
\bottomrule
\end{tabular}
\end{table}

We see three likely reasons for TEN's lead. None has been ablated. First, TEN appears to use a voicing cue. Its source includes a pitch estimator adapted from LPCNet \cite{valin2019lpcnet}. Voiced speech is harmonic. Most LibriVAD noises are not. A periodicity cue therefore stays useful at low SNR, where loudness alone fails. RMS has no such cue. Its absolute dBFS scale also ranks frames poorly once files with different SNRs are pooled. Second, TEN's flat AUC from 20 to 50~ms suggests a fixed internal analysis window. If so, the hop only sets how often a score is produced, not what the model sees. Third, TEN's design matches this benchmark. Its own test set draws on LibriSpeech \cite{tenvad2025}, so read English speech is in its target domain. It is also tuned to catch short pauses, and our forced-alignment labels mark exactly those pauses as silence. Silero's broad multilingual training buys generality, which a single-language read-speech benchmark does not reward.

Two results keep this in proportion. At 30~ms, close to its native chunk, Silero slightly beats TEN (AUC 0.920 versus 0.917). TEN's advantage is therefore about robustness to framing, not about a better model at its best setting. The babble result also supports the voicing view. Babble is itself voiced and harmonic, so a pitch cue gives no separation, and TEN drops to 0.509 AUC at $-5$~dB. Confirming these causes needs per-family results compared against noise harmonicity, out-of-domain data such as VOiCES \cite{richey2018voices}, and a native Silero baseline.

\subsection{Why Silero looks fragile here}
We do not read the Silero curve as evidence that Silero ``prefers'' 30~ms in practice. Our protocol departs from Silero's intended use in four ways. None of them has been ablated, so they should be read as hypotheses.

First, Silero works on fixed chunks. At 20~ms, a 320-sample frame is padded with 192 zeros to reach 512 samples. More than a third of the input is then artificial digital silence. We expect this to push the probability down for real speech frames, which may explain much of the poor AUC at 20~ms. At 30~ms only 32 zeros are added, so the input is close to what the model expects.

Second, for 40 and 50~ms frames we average the probabilities of 512-sample sub-chunks. A 640-sample frame contains one full chunk and one short remainder. The handling of that remainder likely adds noise to the score. Averaging may also blur the sharp transitions that Silero produces.

Third, Silero is a stateful recurrent model. Our script creates one model instance and never resets its state. The state therefore carries across frames within a file and also across file boundaries. When frames are zero-padded, the padding enters the state as if it were real audio. At the start of each file, the state still holds context from the end of the previous, unrelated file, which may bias the first frames.

Fourth, we score raw probabilities. Silero's documented pipeline turns them into speech segments with its own thresholds and duration rules \cite{silero2021}. Our fixed 0.5 threshold skips that step. At 40~ms, F1@0.5 is very low (0.283) while AUC is still 0.803. Ranking therefore survives better than calibration. Scores appear shifted below 0.5, so a fixed threshold misses most speech.

A cleaner protocol would reset state at each file and feed only native 512-sample chunks. It would then map chunk decisions to the reference frames, or compare segments from \texttt{get\_speech\_timestamps()}. Ablations that change one factor at a time (padding, averaging, state reset, post-processing) are needed to confirm which factor drives the drop.

TEN, by contrast, accepts the hop size as a parameter, so no padding or averaging is needed. However, TEN documents only 160 and 256 samples as optimized hops. Its stable results at 320 to 800 samples are an empirical observation, not a supported configuration.

\subsection{Reading the metrics}
\label{sec:metrics_disc}
With a speech prior of 0.745, metrics that reward predicting speech look better than they are. A trivial detector that always says ``speech'' has precision 0.745 and recall 1, so
\begin{equation}
\mathrm{F1}_{\text{trivial}} = \frac{2\cdot 0.745\cdot 1}{0.745+1} \approx 0.854.
\end{equation}
The RMS baseline scores 0.854 to 0.856 F1@0.5, which is no better than this trivial rule. TEN at 0.913 is a real but smaller improvement than the raw number suggests. Similarly, chance-level AP is about 0.745, so an AP of 0.83 under babble at $-5$~dB says little. Table~\ref{tab:deltaf1} makes this explicit by subtracting the trivial F1 from each result. This assumes the prior stays near 0.745 at every window, which holds because all windows share the same sample labels.

\begin{table}[!t]
\caption{F1@0.5 Gain Over a Trivial All-Speech Detector (F1 = 0.854)}
\label{tab:deltaf1}
\centering
\footnotesize
\begin{tabular}{@{}lcccc@{}}
\toprule
Detector & 20 ms & 30 ms & 40 ms & 50 ms \\
\midrule
TEN & $+0.059$ & $+0.059$ & $+0.060$ & $+0.061$ \\
RMS & $+0.002$ & $+0.000$ & $+0.002$ & $+0.001$ \\
Silero & $-0.501$ & $+0.056$ & $-0.571$ & $-0.321$ \\
\bottomrule
\end{tabular}
\end{table}

We therefore treat AUC as the primary metric and recommend that VAD papers on LibriVAD report the speech prior next to AP and F1, as also argued for imbalanced problems in general \cite{saito2015precision}.

\subsection{Small subsets can mislead}
During development we ran a smoke test on the first 10 files. There, TEN scored about 0.48 to 0.49 AUC, RMS about 0.58 and Silero 0.41 to 0.54. Those files happen to be dominated by babble and low SNR, which is the hardest corner of the data. Had we stopped there we would have concluded that TEN is worse than an energy gate. The full run tells the opposite story. We report this because file ordering in generated datasets is often grouped by condition, and quick checks on the first few files are common practice.

\subsection{Implications for our ASR stack}
TEN is already used in one of our two Parakeet servers. On this benchmark it beats the RMS gate by about 0.15 AUC at every window and is far less sensitive to framing than Silero under our protocol. We therefore keep TEN as the default front end. Its quality barely changes between 16 and 50~ms in our sweep. Since only 160 and 256 samples are documented as optimized, we keep the production hop at 256. Babble at low SNR is where TEN struggles most. We did not measure RMS or Silero per condition, so whether they do better there is open. For multi-talker scenes a speaker-aware model is likely needed.

\section{Limitations and Future Work}
This work has clear limits. It is offline and frame level. It says nothing about endpoint latency, early cutoffs or trailing-silence finalization, which shape the user experience in a live system. We evaluated only the small variant and only \texttt{test-clean}. Seen versus unseen noise and out-of-distribution data such as VOiCES \cite{richey2018voices} are not covered. Labels come from forced alignment of clean speech, which can mark short inter-word pauses as silence and may penalize detectors that bridge them. The Silero protocol, as discussed, is not the one its authors intend, and its likely failure causes are not ablated. TEN was run at hop sizes outside its documented range. Only TEN was stratified by noise family and SNR. Finally, all data are English read speech.

Next steps are a streaming endpointing benchmark on the same mixtures, a native 32~ms streaming protocol for Silero, per-stratum plots for RMS and Silero, and runs on the medium and large LibriVAD variants and on non-English data.

\section{Conclusion}
We benchmarked TEN VAD, Silero VAD and an RMS energy baseline on 7.1 hours of noisy LibriVAD speech, at a fixed 16~ms hop and across 20 to 50~ms windows. The main findings are:
\begin{itemize}
\item \textit{TEN is accurate and stable.} It reached 0.917 AUC and 0.968 AP at 16~ms, and its AUC stayed between 0.916 and 0.919 across all windows, including hops outside its documented range.
\item \textit{Energy alone is not enough.} RMS stayed near 0.76 AUC, and its F1 matched a trivial all-speech rule. TEN removes about two thirds of the ranking errors made by RMS.
\item \textit{Silero depends on framing in our protocol.} It equaled TEN near its native 32~ms input but fell to 0.63 AUC at 20~ms. We hypothesize that zero padding, sub-chunk averaging, unreset state and skipped post-processing are likely causes, but these were not ablated. Silero's broad, multilingual design also gets little credit on a single-language read-speech benchmark.
\item \textit{Babble is TEN's main weakness.} At $-5$~dB SNR, TEN is close to chance on babble. The other detectors were not stratified.
\item \textit{Metrics need context.} With a 0.745 speech prior, AP and F1@0.5 overstate quality, so AUC and the prior should be reported together.
\end{itemize}
For a real-time ASR front end that needs flexible framing, TEN is the safer default among the three. The next step is a streaming benchmark that measures endpointing latency and cutoffs on the same data.

\section*{Acknowledgment}
We thank the OneByZero leadership for supporting this work. We also thank the LibriVAD authors for releasing their generator, labels and noise data under an open license.

\bibliographystyle{IEEEtran}
\bibliography{refs}

\end{document}